\documentclass[aps, prd, onecolumn, superscriptaddress, nofootinbib, amsmath, amssymb]{revtex4-2}

\usepackage{graphicx}  
\usepackage{mathrsfs}
\usepackage{amsthm,amsfonts,amsmath}
\usepackage{bbold} 
\usepackage{braket}
\usepackage{physics}
\usepackage[normalem]{ulem}
\usepackage{subcaption}
\usepackage{hhline}
\usepackage{multirow}
\usepackage{xargs} 
\usepackage{tikz}
\usetikzlibrary{shapes.geometric, arrows.meta, positioning, shadows, fit}

\usepackage{todonotes} 

\usepackage[font=small, labelfont=bf, textfont={small,it}]{caption}
\makeatletter

\newcommand{\new}[1]{#1}

\begin{document}

\title{Distributing Quantum Computations, Shot-wise}

\author{Giuseppe~Bisicchia}\email{giuseppe.bisicchia@phd.unipi.it}
\affiliation{Department of Computer Science, University of Pisa, Pisa, Italy}

\author{Giuseppe~Clemente}
\affiliation{Dipartimento di Fisica dell'Universit\`a di Pisa and INFN --- Sezione di Pisa, University of Pisa, Pisa, Italy}

\author{Jose Garcia-Alonso}
\affiliation{University of Extremadura, C\'aceres, Spain}

\author{Juan Manuel Murillo Rodríguez}
\affiliation{University of Extremadura, C\'aceres, Spain}

\author{Massimo~D'Elia}
\affiliation{Dipartimento di Fisica dell'Universit\`a di Pisa and INFN --- Sezione di Pisa, University of Pisa, Pisa, Italy}

\author{Antonio~Brogi}
\affiliation{Department of Computer Science, University of Pisa, Pisa, Italy}

\begin{abstract}
NISQ (Noisy Intermediate-Scale Quantum) era constraints, \new{high} sensitivity to noise and limited qubit count, impose significant barriers on the \new{usability} of QPUs (Quantum Process Units) capabilities. To overcome these challenges, researchers are exploring methods to maximize the utility of existing QPUs despite their limitations. 
Building upon the idea that the execution of a quantum circuit's shots needs not to be treated as a singular monolithic unit, we propose a methodological framework, termed \textit{shot-wise}, which enables the distribution of shots for a single circuit across multiple QPUs. Our framework features customizable policies to adapt to various scenarios. Additionally, it introduces a \textit{calibration} method to pre-evaluate the accuracy and reliability of each QPU's output before the actual distribution process and an \textit{incremental execution} mechanism for dynamically managing the shot allocation and policy updates.
\new{Such an approach enables flexible and fine-grained management of the distribution process, taking into account various user-defined constraints and (contrasting) objectives.}
\new{Experimental findings show that while these strategies generally do not exceed the best individual QPU results, they maintain robustness and align closely with average outcomes. Overall, the shot-wise methodology improves result stability and often outperforms single QPU runs, offering a flexible approach to managing variability in quantum computing.}
\end{abstract}

\maketitle


\section{Introduction}\label{sec:intro}

Advancements in the design and development of Quantum Computers are rapidly accelerating, setting unprecedented milestones and records~\cite{Arute2019,Kim2023,zhou2024electron}. This fast progress brought a plethora of \textit{qubit} implementations and \textit{QPU} (Quantum Process Unit) architectures. Currently, no singular technology reigns supreme, fostering substantial diversity and innovation in quantum computing approaches~\cite{ladd2010quantum}.

However, despite this remarkable technological progress, the capabilities of Quantum Computers remain highly constrained. John Preskill introduced the term \textit{NISQ} (Noise Intermediate-scale Quantum) devices~\cite{preskill2018quantum} to highlight their vulnerability to external noise~\cite{proctor2022measuring} and their limited qubit count, typically ranging from a few dozen to a few hundred~\cite{knill2005quantum}. These constraints severely limit the range of computations feasible within present Quantum Computers, predominantly confining them to tasks requiring only a handful of qubits and a limited number of consecutive operations to mitigate the deleterious effects of noise accumulation~\cite{nisqalgo}.

In response, computer scientists and quantum software engineers are actively engaged in addressing the challenge of maximizing the utility of available quantum devices despite their severe limitations~\cite{nisq}. Their efforts are dedicated to devising strategies that optimize quantum computations within current technological boundaries.
To this end, researchers are designing increasingly sophisticated techniques for quantum error detection~\cite{andersen2020repeated}, mitigation~\cite{endo2018practical}, and correction~\cite{lidar2013quantum}. Moreover, strategies for ``cutting'' quantum circuits that exceed the capacity of NISQ devices are currently under active investigation~\cite{peng2020simulating,cutqc}. Other approaches involve the careful selection of the most appropriate quantum computer for each computation, taking into account performance metrics and limitations of the available options as well as the characteristics of each specific quantum circuit~\cite{qapigateway,nisqanalyzer,Bisicchia2023363}. 

In this paper, we introduce an innovative approach to face the constraints of current NISQ devices and to \new{improve} the effectiveness of quantum computations. Our methodology diverges from conventional strategies by proposing a shift in perspective regarding the execution of quantum tasks. Traditionally, executing a quantum circuit typically consists of executing iteratively numerous independent times, referred to as ``\textit{shots}'', due to the inherently probabilistic nature of quantum mechanics and qubits. Consequently, the output of a quantum computation usually does not consist of a single measured state from a single run. Instead, it includes a distribution reflecting the frequency of output states obtained after running the quantum circuit for multiple (usually thousands) shots (i.e., iterations).

In our approach, we propose a departure from viewing the execution of the shots of a quantum circuit as a single, indivisible unit that must be completed in a solitary run. Instead, we advocate for a more flexible and fine-grained perspective, offering various degrees of freedom in the process. Specifically, we suggest that even for a single circuit, its shots can be distributed, or ``\textit{split}'', across multiple heterogeneous Quantum Computers based on specific \textit{custom} policies~\footnote{\new{Note that our proposed approach differs from shot-reduction optimization strategies such as the one in~\cite{zhu2024optimizing}. Indeed, our focus is on distributing a particular amount of shots on multiple, independent Quantum Computers. Still, as we will discuss in Sect.~\ref{sec:methods}, our framework is also capable of reducing the total amount of shots performed.}}. Subsequently, the results obtained from each Quantum Computer, representing the output distributions of the circuit execution for that particular QPU, are then merged together in a unified output distribution. \new{In the rest of this paper we both discuss the general methodological framework and propose several strategies to \textit{split} and \textit{merge} the shots of a quantum circuit (e.g., by distributing the shots equally to each Quantum Computer, in a random way or proportionally to the estimated ``reliability'' of each QPU).}


\new{Through this approach, we aim to leverage the limitations of the NISQ era and turn them into advantages. In the experimental section of this paper, we present findings on the execution of quantum circuits using multiple QPUs, focusing on the impact of split and merge strategies. The experiments encompass various circuit types and assess performance against established baselines obtained from single QPU runs. Key findings indicate that while split-merged strategies do not consistently exceed the best baseline performance, they maintain robustness and are generally aligned with average baseline outcomes. Notably, performance disparities arise across different circuits, suggesting that circuit-specific characteristics influence results. These insights lay the groundwork for exploring advanced strategies in quantum computing.}

\new{Furthermore, as discussed in our previous work~\cite{bisicchia2023dispatching} and further elaborated in~\cite{Bisicchia2023363}, adopting a shot-wise management approach to quantum computation offers various qualitative advantages. These encompass enhanced fault resilience to QPU failures, finer-grained management, greater customizability to user requirements, and reduction of waiting times. With respect to our preliminary work~\cite{bisicchia2023dispatching,Bisicchia2023363}, in this article, we define, formalize, and improve the shot-wise distribution methodology with a more general and holistic approach to the problem and perform numerical experiments to assess the methodology distributing the shots up to seven QPUs from two different manufacturers and two different qubit implementations. Moreover, we discuss and develop four distribution policies, two of them informed by the expected noise of each QPU.}

\new{Experimental results show that by enabling the distribution of shots across multiple NISQ devices, our proposal makes it possible to reconcile multiple conflicting objectives (such as waiting time, price and reliability of a quantum computation~\cite{Bisicchia2023363,bisicchia2023dispatching}). Moreover, distributing and merging the shots of a particular quantum circuit on various noisy QPUs produces final output distributions more reliable than performing the whole computation on a single QPU (as discussed in the experimental section of this paper, Sect.~\ref{sec:numres}). As a final advantage, the framework offers high flexibility in the policies and strategies to assess the capabilities of the available QPUs, how to distribute and merge the shots, and how to dynamically optimize such procedures and reduce the number of shots. To the best of our knowledge, ours is the first work proposing a methodology that enables all the above features and capabilities.}

\new{Summarizing, the main contributions of this paper are:}
\begin{enumerate}
    \item[\new{(a)}] \new{the proposal of an innovative, general framework to distribute the shots of a single quantum circuit on multiple NISQ devices while also estimating the ``reliability'' and dynamically optimizing the shots allocation,}
    \item[\new{(b)}] \new{the definition of various policies to \textit{split} and \textit{merge} the shots, and}
    \item[\new{(c)}] \new{an experimental evaluation of our proposed approach on multiple quantum circuits, providers and QPUs.}
\end{enumerate}

This paper addresses the motivation and discussion surrounding the shot-wise distribution of quantum computations across heterogeneous Quantum Computers. We present a comprehensive methodology framework to implement such a distribution strategy, parameterized by a set of customizable policies. We illustrate the various degrees of freedom and the underlying criteria guiding each possible decision (Sect.~\ref{sec:general}). Subsequently, we discuss various potential split and merge policies (Sect.~\ref{subsec:policies}). We then detail the experimental protocol devised to evaluate and validate the shot-wise approach, and we analyze the results obtained from these experiments (Sect.~\ref{sec:numres}). Finally, we provide an overview of related work (Sect.~\ref{sec:literature}) and draw some conclusions, also mentioning potential directions for future research (Sect.~\ref{sec:concl}).

\section{General Framework}\label{sec:methods}

In this Section, we introduce some definitions and fix the notation used in the following discussions.
\new{An overview of the main concepts is provided in standard book references such as~\cite{Nielsen2012}.}

\subsection{Preliminaries}

Let us consider a circuit $U\in SU(2^q)$ (for which we extensively use
the equivalence between ideal circuits and unitary operators) acting on $q$ qubits initialized
as $\ket{0}$. In the ideal noiseless case, the output of 
a single execution is a pure state $\ket{\psi}=U\ket{0}$.
A measurement in the computational basis 
$\{\ket{x}\}_{x\in \mathbb{Z}_{2^q}}$, will yield a specific bitstring $x$
with probability $p^{(\text{ideal})}_{x}={|\mel{x}{U}{0}|}^2$.
Any other initialization and change of basis for the measurement 
can be incorporated into the circuits without loss of generality.
Moreover, in the presence of quantum noise, 
the final state can be represented as a density matrix $\rho$, 
obtained from the application of a noisy quantum channel $\mathcal{E}$ to the initial standard pure state $\rho_0=\ketbra{0}{0}$ 
as $\rho = \mathcal{E}(\rho_0)$.
Therefore, the probability of measuring the bitstring $x$ in the computational basis from a single execution is distributed according to a discrete probability $p_x= \Tr[\ketbra{x}{x}\rho]$, where $\rho=\mathcal{E}_{U}(\ketbra{0}{0})$ depends on the quantum channel $\mathcal{E}_{U}$, which only approximates the effect of the ideally unitary circuit that implements $U$\footnote{In the ideal case of a unitary channel, one would have $\mathcal{E}_{U}(\rho_0) = U \rho_0 U^\dag$, while in general one has $\mathcal{E}_{U}(\rho_0)=\sum_{\alpha} K_\alpha \rho_0 K_\alpha^\dag$ in the Kraus representation~\cite{Nielsen2012}.}. Therefore, for any fixed circuit $U$ and QPU $m$, 
we expect the biases $p_x-p^{(\text{ideal})}_{x}$ to be in general non-vanishing,
signaling a discrepancy in the results even in the limit of unlimited resources,
unless an error correction or mitigation scheme is applied.
Repeating the experiment a number $n$ of times, commonly known as \emph{shots},
the number of times a specific state labeled $x$ is measured is
called \emph{counts} and denoted by $\hat{c}_x$, 
and it follows a multinomial distribution based on $p_x$, i.e.,
\begin{align}\label{eq:multinomialP}
     P\Big(\hat{c}_x=c_x \forall x=0,\dots,2^q-1; \sum_x c_x=n\Big) 
     \equiv n! \prod_{x\in \mathbb{Z}_{2^q}} \frac{{(p_x)}^{c_x}}{c_x!}.
\end{align}
From the relative counts, it is possible to estimate 
the underlying probability distribution as 
$\hat{p}_x \equiv \frac{\hat{c}_x}{n}$,
which is unbiased ($\mathbb{E}[\hat{p}_x]=p_x$) and
fluctuates with a statistical error estimated as $\Delta p_x=\sqrt{\frac{1}{(n-1)}\hat{p}_x (1-\hat{p}_x)}$.

We define a \emph{$q$-quantum processing unit} ($q$-QPU) as a hardware
capable of running a generic quantum circuit with $q$ qubits.
Notice that, with this definition, different connected subtopologies of
$q<K$ qubits on the same $K$-QPU are treated as different $q$-QPUs. 
This allows to include the possibility of considering disconnected subtopologies
of the same QPU (which might make sense only if the crosstalk between the subtopologies involved is negligible). 

We define a \emph{policy} as a set of criteria determining the specific decisions taken at different
steps of the heterogeneous quantum computation.
This can involve, for example, limitations and prior knowledge about a QPU, different kinds of budget constraints, time constraints, 
optimization methods, and whatever the arbitrary choices of the user are.
In the following discussions, we denote any policy 
by the symbol $\mathscr{P}$. 
In the next Section, we propose a framework
of heterogeneous quantum computation where
the set of policies acts as a controller.
Different possible policy choices are discussed thoroughly in Section~\ref{subsec:policies}. 

\subsection{General strategy for heterogeneous quantum computation}
\label{sec:general}
In this work, the general aim of an optimal heterogeneous computation involves allocating computational resources (\textit{split}), gathering and merging results (\textit{merge}), 
and updating the information on the performance with optimality criteria, 
according to the policies chosen by the user\footnote{\new{Note that the choice of the split strategy and the merge policy are completely independent.}}.
However, for a starting set of QPUs, prior information on performance and accuracy of each QPU are not always available or comparable to each other.
This lack of information makes the merge inaccurate in some cases,
for example, when the results of the majority of Qs cluster around 
some point in probability space, while the most accurate QPUs lie far from the majority and would be treated as outliers. 

This situation can be overcome by a \emph{Calibration} 
and \emph{Ranking} stage, which would then guide further updates 
during the \emph{Production} \new{(i.e., execution)} stages.
Anyway, whenever available and comparable, one can rely directly also on external calibration data from quantum providers
for specific quantum devices.
Otherwise, we propose a calibration scheme tailored to the 
specific device subtopologies and the class of circuits one aims to execute 
in production, provided it is possible to compute ideal probability distributions
to be used as benchmarks against which one can compare noisy results from real QPUs. Of course, this computation is possible on classical computers through
emulation only for a limited number of qubits. The calibration stage
is instrumental for a ranking of the QPUs considered since it allows
to associate an \emph{unreliability index} to each QPU.
After these stages, which can be performed with a fraction of the full budget
expected for the full run, one can proceed with the production stages, which
involve an iterative scheduling of split-merge-update steps
where the basic execution task involves measurements on some target circuits 
but without any ideal benchmark. 
The split part on the very first iteration is determined by the calibration 
and ranking stages, if available, while the merge part depends on all the previous stages, as well as on the results of the last execution. 
Finally, the unreliability information can be updated at the end of each production stage based on the data accumulated at each iteration. 
One might also consider embedding the production stage into a pipeline where QPU executions happen asynchronously. In this case, one can continuously update the information, merging data according to the available partial information and proceeding with the next steps.  
Furthermore, a stopping policy can be considered, which allows an earlier termination of the run in case a stopping criterion is reached.

\begin{figure}
    \centering

\begin{tikzpicture}[node distance=0.5cm and 4.0cm,
    execution/.style={rectangle, text width=5cm, align=flush left, draw=black, fill=red!15, drop shadow},
    process/.style={rectangle, rounded corners, text width=5cm, align=flush left, draw=black, fill=blue!10, drop shadow},
    end/.style={rectangle, rounded corners, text width=5cm, align=flush left, draw=black, fill=orange!20, drop shadow},
    decision/.style={diamond,aspect=4.5,inner sep=1pt,text width=5cm, align=center, draw=black, fill=green!15, drop shadow},
    arrow/.style={-Latex}]


\begin{scope}[local bounding box=calibration]
\node (chooseQMs) [process] {(C.1) Choose a prior shot allocation $n_m^{(0)}$ for each QPU $m$};
\node (chooseCircuits) [process, below=of chooseQMs] {(C.2) Choose a set of circuits $\{U_c\}$ and compute $p_{x}^{(\text{ideal},c)}$};
\node (executeCircuits) [execution, below=of chooseCircuits] {(C.3) Execute each circuit $U_c$ on each QPU $m$, obtaining $\hat{c}_x^{(m,c)}$};
\end{scope}

\begin{scope}[local bounding box=ranking,below=of calibration]
\node (assignUnreliability) [process,below=of executeCircuits,yshift=-1.5cm] {(R) Estimate unreliability $u_m$ for each QPU $m$};
\end{scope}

\begin{scope}[local bounding box=production, right=of calibration,xshift=5.7cm,yshift=0.2cm]
\node (computeSplitCalib) [process] {(P.0) Compute preliminary split weight $w_m^{(1)}$};
\node (prodSched) [process, below=of computeSplitCalib] {(P.1-\textit{i}) Allocate resources for the scheduled $\textit{i}$-th iteration $n_m^{(i)}$};
\node (executeU) [execution, below=of prodSched] {(P.2-\textit{i}) Execute the circuit $U$ on each QPU $m$};
\node (mergeDistributions) [process, below=of executeU] {(P.3-\textit{i}) Perform merge policy of the partial distributions};
\node (improveSplit) [process, below=of mergeDistributions] {(P.4-\textit{i}) Update split weights for the next iteration};
\node (checkStop) [decision, below=of improveSplit] {(P.5-\textit{i}) Check stop condition};
\end{scope}

\node (stop) [end, below=of checkStop,align=center,yshift=-0.7cm] {Terminate};

\draw [arrow] (chooseQMs) -- (chooseCircuits);
\draw [arrow] (chooseCircuits) -- (executeCircuits);

\draw [arrow] (computeSplitCalib) -- (prodSched);
\draw [arrow] (prodSched) -- (executeU);
\draw [arrow] (executeU) -- (mergeDistributions);
\draw [arrow] (mergeDistributions) -- (improveSplit);
\draw [arrow] (improveSplit) -- (checkStop);
\draw [arrow,line width=1.5pt] (checkStop) -- node[anchor=east,yshift=-0.3cm] {yes} (stop);
\draw [arrow,line width=1.5pt] (checkStop.east) -- ++(1,0) |- node[near start, anchor=south west] {no} (prodSched.east);

\node (calibrationBox) [draw=black!50, dashed, rounded corners, fit=(calibration), inner sep=6mm] {};
\node[fill=white] at (calibrationBox.north) {Calibration Stage};
\node (rankingBox) [draw=black!50, dashed, rounded corners, fit=(ranking), inner sep=4mm] {};
\node[fill=white] at (rankingBox.north) {Ranking};
\node (productionBox) [draw=black!50, dashed, rounded corners, fit=(production), inner sep=4mm] {};
\node[fill=white] at (productionBox.north) {Production Stage (target circuit $U$)};

\draw [arrow, line width=1.5pt] (calibrationBox.south) -- ([shift=({0,0.20})]rankingBox.north) ;
\draw [arrow, dashed, line width=1.5pt] (calibrationBox.east) -- ++(0.5,0) |- (productionBox.west);
\draw [arrow, dashed,line width=1.5pt] (rankingBox.east) -- ++(0.7,0) |- (productionBox.west);

\end{tikzpicture}
    \caption{Diagram of the general strategy of heterogeneous quantum computation 
    discussed in the text.
    Thin arrows connect steps
    in sequence, while thick arrows describe
    dependencies (with dashed line describing optional dependency).}
    \label{fig:diagram_strategy}
\end{figure}
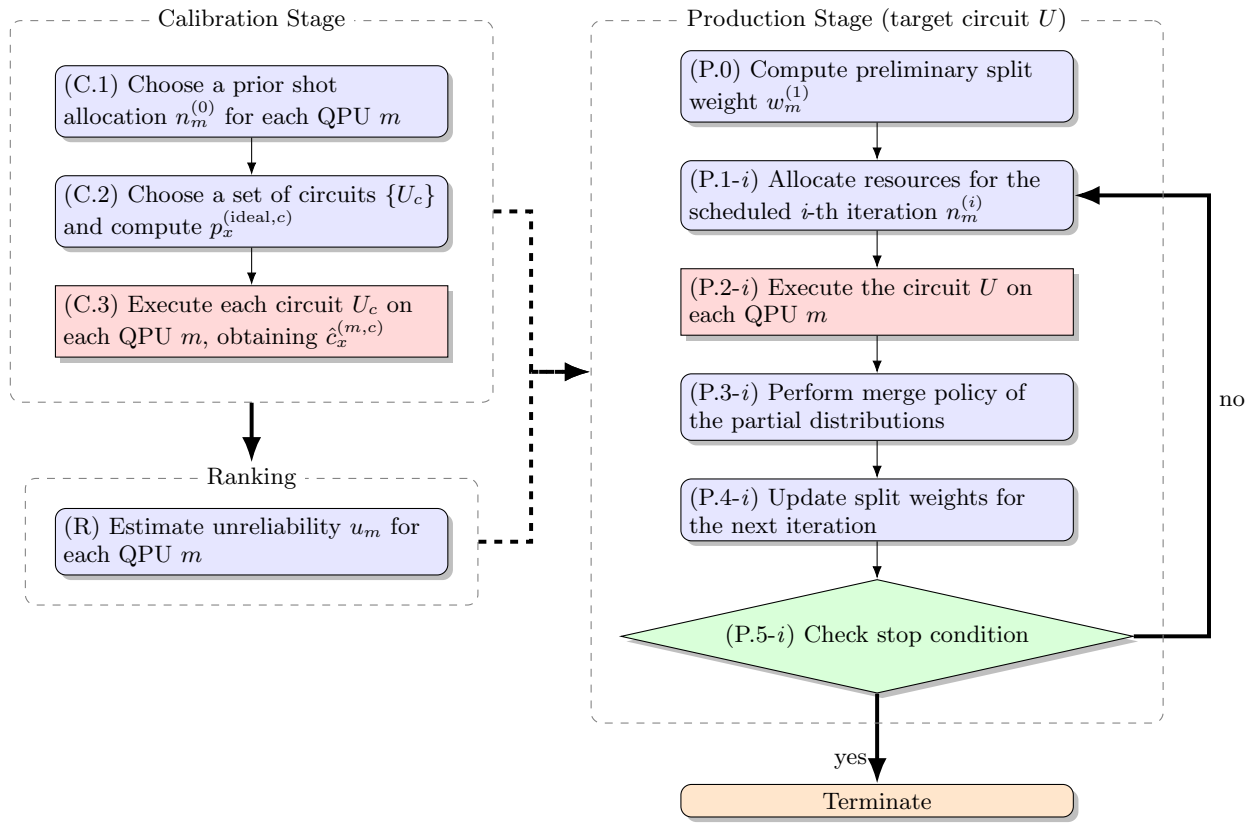

Having outlined the general strategy of heterogeneous computation we propose, here it follows a scheme for a given set $\mathcal{M}$ of $M\equiv |\mathcal{M}|$ $q$-QPUs, while in Fig.~\ref{fig:diagram_strategy} we display a diagrammatic depiction of the processes involved.\\

Calibration and Ranking stages:
\begin{enumerate}
    \item[(C.1)] \textbf{$\mathscr{P}^{(\text{calib})}_{\text{prior-split}}$ --- initial split:} choose a prior shot allocation $\vec{n}^{(0)} \equiv (n_m)_{m\in \mathcal{M}}$ (for example, fixing the total number of shots $n_{\text{tot}}^{(0)}=\sum_m n_m^{(0)}$ and using a uniform allocation $n_m^{(0)} = n_{\text{tot}}^{(0)}/M$);
    \item [(C.2)] \textbf{$\mathscr{P}^{(\text{calib})}_{\text{bench}}$ --- benchmark:} choose a ``training set'' of circuits $\mathcal{C}$, represented ideally by a class of unitary operators $\{U_c\}_{c\in \mathcal{C}}$ acting on $q$ qubits, and compute the ideal probability distribution of measurements in the computational basis, denoted as $p_{x}^{(\text{ideal},c)}\equiv |\mel{x}{U_c}{0}|^2$. These will be used as benchmark distributions;
    \item[(C.3)] \textbf{Calibration executions:} execute each circuit $U_c$ on each QPU $m\in \mathcal{M}$ (starting conventionally from the pure state $\ket{0}$) with 
    the selected number of shots $n_m^{(0)}$, obtaining the counts $\hat{c}_{x,c}^{(m)}$, distributed according to Eq.~\eqref{eq:multinomialP} with a generating probability given by the specific noisy (and unknown) realization $p_x^{(m;c)}$;
    \item[(R)] \textbf{$\mathscr{P}^{(\text{calib})}_{\text{rel}}$ --- unreliability:}
    using the results from the previous step and comparing them
    with the benchmark distributions,
    assign an \emph{unrealiability} coefficient $u_m$ to each QPU $m$ 
    and compute the optimal split policy for the next stage
    in the form of a proposal for shot split weight $w_m^{(1)}$;
\end{enumerate}
Production stage (starts with $\textit{i}=1$, target circuit $U$):
\begin{enumerate}
    \item[(P.0)] \textbf{$\mathscr{P}^{(\text{prod})}_{\text{init}}$ --- calibrated prior split weights:} 
    using the results from the calibration step (C.3), and comparing them
    with the benchmark distributions,
    compute the optimal split policy for the next stage
    in the form of a proposal for shot split weight $w_m^{(1)}$;
    \item[(P.1-\textit{i})] \textbf{$\mathscr{P}^{(\text{prod})}_{\text{split}}$ --- production split:} 
        implements \textit{i}-th iteration of the production schedule, computing the number of shots to run in this iteration (for example, simply dividing the total number of shots by the the maximum number of iterations) and an
        optimal split according to prior information;
    \item[(P.2-\textit{i})] \textbf{Production executions:}
    execute the circuit $U$ on each QPU $m$, using a given the number of shots of the current interaction and the split policy, by allocating shots according to the split weights $w_m^{(\textit{i})}$ computed in the previous iteration of the production stage (P.4-(\textit{i}-1)), or in the first step of the production stage (P.0) if this was the first iteration (i.e., $\textit{i}=1$);
    \item[(P.3-\textit{i})] \textbf{$\mathscr{P}^{(\text{prod})}_{\text{merge}}$ --- merge results:}
    perform the merge policy of the partial distributions estimated by the execution after split. This might also include some error-mitigation strategy per-QPU and per-circuit;
    \item[(P.4-\textit{i})] \textbf{$\mathscr{P}^{(\text{prod})}_{\text{update}}$ --- update optimal split:}
    improve the prior split weights for the next step by analyzing the relative performance of different QPUs by accumulating
     statistics from all previous runs of $U$ and proposing a new split weight
    $w_m^{(\textit{i}+1)}$ and the number of shots to perform in the next iteration;
    \item[(P.5-\textit{i})] \textbf{$\mathscr{P}^{(\text{prod})}_{\text{stop}}$ --- check stopping criterion:} 
    if the conditions for the chosen stopping policy are not met, proceed with point (P.2-(\textit{i}+1)), otherwise terminate.
\end{enumerate}
In the next section, we mention some of different possible choices for the policies to be used in the calibration and production stages.

\subsection{Policies}\label{subsec:policies}
\subsubsection{Calibration --- initial split}\label{subsubsec:poly-calib-initsplit}
Deviations from a uniform initial split at calibration can be motivated by various factors: the economic cost per shot for each QPU, different queue and execution times, some information about the accuracy of each QPU for the task in question. In general, this policy should reflect all the preferences of the user regarding all or some of these aspects (and possibly others), so that the actual split used can be optimized by considering both information about the QPUs and the specific user needs. 
\subsubsection{Calibration --- benchmarks}
Depending on the task considered in production, it might be possible to identify a class of circuits that can be used as a ``training'' set for the calibration stage, so that one can provide with more tailored information for the production stage. Even if the tasks considered are generic, it could be useful, for example, to test the QPUs on a set of random circuits. This gives just a crude estimate of the general performance of the QPU but, as we show in Section~\ref{sec:numres}, the performance at calibration on a fixed set of random circuits does not necessarily reflect the performance observed on specific tasks.
Notice also that the results of the executions for each 
of these benchmarking circuits should be compared with exact results, which means that this step is accessible only for circuits involving a relatively small number of qubits so that one can perform an exact emulator as a benchmark, or at least with an output distribution that is simple enough to be easily computable.
\subsubsection{Calibration --- executions}
The benchmark circuits are executed on each QPU with the selected number of shots.
Some preprocessing and postprocessing might be involved at this step.
For example, the circuit might be decomposed in different primitive gates for each QPU  or one could apply different mitigation strategies (provided this enters the shot budget or other resource criteria). Since the calibration stage is preparatory to
the actual production stage, the same techniques are expected to be applied 
also later during the production stage executions.
\subsubsection{Ranking --- unreliability}
As a result of the calibration, we want to assign a ``goodness'' value to each QPU considered, so that the production stage can be guided by it.
The specific metric can also depend on different factors, but it should reflect the discrepancy 
between the results of the QPUs on the set of benchmarking circuits
and the exact output distributions. A distance between probability distributions (or density matrices) is, therefore, an essential component of this policy. 
For example, in Section~\ref{subsec:numres-calib} we adopt
as the ``unreliability'' factor the Hellinger distance between the results of QPU executions and the exact one, averaged among all the benchmarking circuits. 
\subsubsection{Production --- prior split weights}
For this policy, the same aspects mentioned in~\ref{subsubsec:poly-calib-initsplit} can be considered. Furthermore, the results of the calibration stage, if available, can be integrated into the analysis as expected prior accuracy provided by each QPU.
We reason here in terms of prior split ``weights'' because the specific shot allocation might change depending on more information from the production schedule.
\subsubsection{Production --- split strategies}
In the single iteration version of the production stage, this step is a trivial application of the prior split weights step mentioned above applied to the total number of shots. In the case where more iterations of the production stage loop are needed, different shot allocations might be involved based on an update of the prior collective information depending on former results and a different amount of shots can be considered at each iteration.
Besides schedule-dependent choices, one can test different policies that use the prior information in different ways. 
In this work, we investigate three specific cases of splitting strategy: 
\begin{itemize}
    \item \emph{uniform}: the total number of shots for that iteration are allocated uniformly
            among all the QPUs;
    \item \emph{Hellinger}: the shots are allocated depending on the split weights computed using the Hellinger distance (discussed in Section~\ref{subsec:poli-distr-dists}) between the count distributions and either the exact one, if provided at calibration, or the best expected one estimated from previous observations during the production stage;
    \item \emph{MISE}: in this approach, referred to \emph{Mean Integrated Square Error} (see Section~\ref{subsubsec:OptMISE}), the shots are allocated in such a way as to minimize not only 
    the relative discrepancy from the exact distribution at calibration (or the best expected one from previous iterations of the production stage), taking into account also the expected statistical error coming from a finite number of shots per QPU.
\end{itemize}
\subsubsection{Production --- executions}
The same considerations done during calibration executions apply here.
\subsubsection{Production --- merge strategies}
In this step, after gathering all counts obtained from the executions on each QPUs, one has to merge the results. As for the split strategies discussed above, one can take into consideration different factors involved, but many, such as shot cost or queue time should not play a role, since the data is already assumed to be fully available after the executions (or, in the case of a pipelined production stage, of the partial information available).
The main goal of this step is therefore to maximize the \emph{accuracy} of the final result with the given information on the executions. 
In analogy with the split strategies, in this work, we investigate three specific cases: 
\begin{itemize}
    \item \emph{uniform}: the distributions estimated from each QPU are simply merged (summing all counts per each outcome);
    \item \emph{Hellinger}: the target distribution is chosen as the one that minimizes the total squared Hellinger distance (discussed in Section~\ref{subsubsec:OptHSD}) weighted according to unreliabilities and other supplementary information;
    \item \emph{MISE}: in this case, the target distribution is chosen as the optimal convex sum of all the QPU distributions, where optimality is determined as the minimum of the mean integrated square error (MISE), which takes into account both bias and variance of the data, as described in Section~\ref{subsubsec:OptMISE}.
\end{itemize}

\subsubsection{Production --- update split}
This step is required in the case of a schedule with more than a single iteration 
since it involves the updating of both the number of shots to split and the prior split weights which would be used as improved collective information at the beginning of the next iteration of the production loop. 
\subsubsection{Production --- stopping criterion}
In the cases when one decides to perform more than once the steps in the production stage, different choices of the
stopping criterion might be preferred. For example, a straightforward stopping policy might just be the depletion of the total shot (or any other kind of) budget cap expected 
or the reach of some accuracy and/or precision requirement on the final result (which might result in a lower total budget expense and faster global execution). 

\subsection{Distance between discrete probability distributions}\label{subsec:poli-distr-dists}
For the following discussions, 
it is useful to introduce a metric of comparison in the form of a distance between probability distributions,  which are considered here as the main output of the execution of a quantum circuit, as measurements in the computational basis.
In a general quantum setting, one is interested in the distance between density matrices.
Nevertheless, for a wide class of quantum algorithms (e.g. Grover searches, some measurements in real-time evolution),  the output of a quantum circuit is a collection of measurements on a fixed basis, while a change of basis can usually be incorporated into the circuit, after which
measurements in the computational basis follow.
Even if multiple changes of basis are needed (for example, for a typical VQE algorithm), one can consider each version as a distinct circuit,
to which the whole analysis can be independently applied.
Therefore, it is possible to define as a single \emph{task} a collection of measurements on a fixed circuit in the computational basis, so that a probability distribution of outcomes can be inferred from the relative counts, while more complex algorithms involve more than
one of these simple tasks in general.

We consider the Hellinger distance~\cite{Hellinger_original} 
between two discrete distributions $p_x$, $q_x$, defined as
\begin{align}\label{eq:Hellinger_def}
   d_{H}(p,q) \equiv \sqrt{\frac{1}{2}\sum_{x} {(\sqrt{p_x} - \sqrt{q_x})}^2} 
   = \sqrt{1-\sum_{x} \sqrt{p_x q_x}}=\sqrt{1-\cos \Delta(p,q)},
\end{align}
where in the rightmost term we introduced the so-called \emph{Bhattacharyya angle}~\cite{bhattacharyya1943measure}, defined as the angle
$\Delta(p,q) = \arccos{\sum_{x} \sqrt{p_x q_x}}$ between the vectors $(\sqrt{p_x})$ and $(\sqrt{q_x})$, both with Euclidean ($l^2$) norm 1 and with positive components.
In our case, we do not have direct access to the probability distributions
$p^{(m)}_x$ for each QPU $m$;  only the information about the counts $c_x^{(m)}$, obtained using a finite number of shots $n^{(m)}$, is available.
Due to the non-linearity of the definition 
of the Hellinger distance in Eq.~\eqref{eq:Hellinger_def}, 
replacing the best estimate $\hat{p}_x^{(m)}=\frac{\hat{c}_x^{(m)}}{n^{(m)}}$
typically yields a biased estimate, 
namely\footnote{By Jensen inequality, 
$\mathbb{E}[f(\hat{X})]\geq f(\mathbb{E}[\hat{X}])$ 
for a convex function $f$ of a random variable $\hat{X}$; 
the opposite inequality is true for a concave function 
such as $x\mapsto \sqrt{x}$. Therefore 
$\mathbb{E}[{d_H(\hat{p},q)}^2]=1-\sum_x \mathbb{E}[\sqrt{\hat{p}_x}] \sqrt{q_x} \geq 1-\sum_x \sqrt{p_x} \sqrt{q_x} = {d_H(p,q)}^2$.},  $\mathbb{E}[d_H^2(\hat{p},\hat{q})]\geq \mathbb{E}[d_H^2(\hat{p},q)]\geq d_H^2(p,q)$.
Due to this bias, to properly estimate these distances, 
one has to apply a statistical technique of bias removal
as outlined in Appendix~\ref{sec:biasremoval}.

\subsubsection{Weighted average square Hellinger distance}\label{subsubsec:OptHSD}
Using the distance metric discussed in the previous section, we can estimate the difference between the relative counts for the dataset $\mathcal{D}^{(m)}$ and the ideal target distribution $p^{(\text{ideal})}_x$
known at calibration stage, 
where the bias and associated errors are estimated with the techniques discussed in Appendix~\ref{sec:biasremoval}.
These distances can then be used as \emph{unreliability}
parameter to be associated with each QPU.
A pre-ranking of the QPUs can be determined by reordering them from the lowest unreliability to the highest.
We consider an optimal probability distribution $\bar{p}_x$ as the one 
that minimizes the weighted average square distance with respect to 
the distributions estimated from the QPU results, 
namely 
\begin{align}\label{eq:DSquare}
D^2(\bar{p};p^{(m)},w^{(m)}) = \sum_{m=0}^{M-1} w^{(m)} d^2(\bar{p},p^{(m)}),
\end{align}
where $w^{(m)}$ are the \emph{reliability weights} associated to each QPU.
\new{In the cases where $d$ corresponds to the Hellinger distance, the probability distribution $\bar{p}^*$ which minimizes $D^2$ is described in Appendix~\ref{sec:WSHD_weights_opt}.}

\subsubsection{Mean Integrated Square Error}\label{subsubsec:OptMISE}
Let us consider a single-circuit benchmark 
with ideal distribution $p^{(\text{ideal})}_x$, 
and collection of $M$ QPUs, with distribution
$p^{(m)}_x$, sampled with a certain number of shots 
$n_m$, depending on the split policy 
and compactly denoted by the ``split-shot'' vector \mbox{$\vec{n}=(n_m)$}.
Any convex merge policy is defined by a weight
vector $\vec{w}={(w_m)}_{m=0}^{M-1}$ and a
weighted distribution estimator as follows
\begin{align}\label{eq:counted_distr_weighted}
    \hat{p}^{(\vec{w};\vec{n})}_x \equiv \sum_{m=0}^{M-1} w_m \hat{p}^{(m;n_m)}_x, \quad \text{where}\quad
    \hat{p}^{(m;n_m)}_x\equiv \frac{1}{n_m}\sum_{y\in\mathcal{D}^{(m)}} \delta_{x,y}.
\end{align}
The variables in Eq.~\eqref{eq:counted_distr_weighted}
are unbiased estimators of
which is an unbiased estimator of $p^{(\vec{w})}_x \equiv \sum_{m=0}^{M-1} w_m p^{(m)}_x$,
which we want to make as close as possible 
to $p^{(ideal)}_x$ by optimizing 
the weight parameters $\vec{w}$.
Since each QPU contributes in general 
with a different number of shots, a dataset realization $\mathcal{D}$ can be formally decomposed into independent sub-datasets $\mathcal{D}=\cup_{m=0}^{M-1} \mathcal{D}^{(m)}$ 
such that $|\mathcal{D}^{(m)}| = n_m$ with 
$\mathcal{D}^{(m)}$ sampled according to a multinomial distribution with $p_x^{(m)}$ as probability for each extraction $x$.
In the following discussion we consider the Mean Integrated Square Error, defined as
\begin{align}\label{eq:MISEdef}
\textrm{MISE}(\vec{w};\vec{n}) \equiv \mathbb{E}_{\mathcal{D}=\cup_{m} \mathcal{D}^{(m)}} \Big[\sum_x {(\hat{p}_x^{(\vec{w};\vec{n})}[\mathcal{D}] - p^{(\text{ideal})}_{x})}^2 \Big], 
\end{align}
where the expectation value involves all possible 
realizations of the full dataset $\mathcal{D}$ with
fixed split-shots $\vec{n}$ and according to 
their probabilities.
\new{More details on the MISE definition and optimization are reported in Appendix~\ref{sec:MISE_opt}.}

\section{Experimental Assessment}\label{sec:numres}
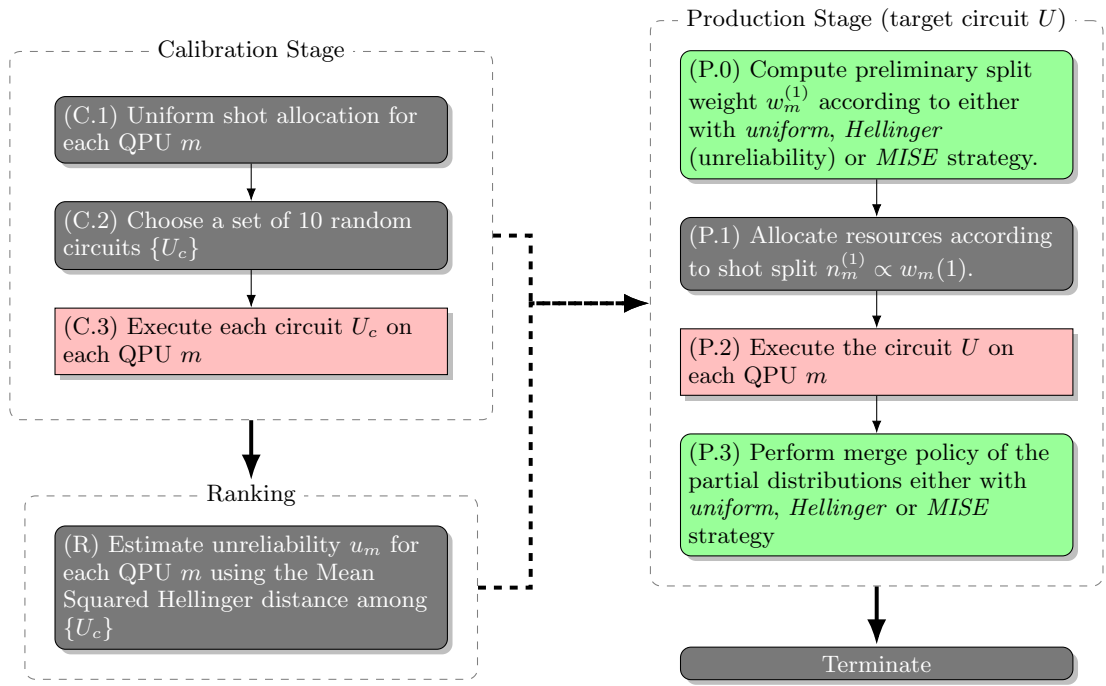
\begin{figure}
    \centering
\begin{tikzpicture}[node distance=0.5cm and 4.0cm,
    execution/.style={rectangle, text width=5cm, align=flush left, draw=black, fill=red!25, drop shadow},
    process/.style={rectangle, rounded corners, text width=5cm, align=flush left, draw=black, fill=green!40, drop shadow},
    process_fixed/.style={rectangle, rounded corners, text=white,text width=5cm, align=flush left, draw=black, fill=darkgray!70, drop shadow},
    end/.style={rectangle, rounded corners, text width=5cm, align=flush left, draw=black, text=white, fill=darkgray!70, drop shadow},
    decision/.style={diamond,aspect=4.5,inner sep=1pt,text width=5cm, align=center, text=white, draw=black, fill=darkgray!70, drop shadow},
    arrow/.style={-Latex}]

\begin{scope}[local bounding box=calibration]
\node (chooseQMs) [process_fixed] {(C.1) Uniform shot allocation for each QPU $m$};
\node (chooseCircuits) [process_fixed, below=of chooseQMs] {(C.2) Choose a set of 10 random circuits  $\{U_c\}$};
\node (executeCircuits) [execution, below=of chooseCircuits] {(C.3) Execute each circuit $U_c$ on each QPU $m$
};
\end{scope}

\begin{scope}[local bounding box=ranking,below=of calibration]
\node (assignUnreliability) [process_fixed,below=of executeCircuits,yshift=-1.5cm] {(R) Estimate unreliability $u_m$ for each QPU $m$ using the Mean Squared Hellinger distance among $\{U_c\}$};
\end{scope}

\begin{scope}[local bounding box=production, right=of calibration,xshift=5.7cm,yshift=0.2cm]
\node (computeSplitCalib) [process] {(P.0) Compute preliminary split weight $w_m^{(1)}$ according to either with \emph{uniform}, \emph{Hellinger} (unreliability) or \emph{MISE} strategy.};
\node (prodSched) [process_fixed, below=of computeSplitCalib] {(P.1) Allocate resources according to shot split $n_m^{(1)}\propto w_m{(1)}$.};
\node (executeU) [execution, below=of prodSched] {(P.2) Execute the circuit $U$ on each QPU $m$};
\node (mergeDistributions) [process, below=of executeU] {(P.3) Perform merge policy of the partial distributions 
either with \emph{uniform}, \emph{Hellinger} or \emph{MISE} strategy};
\end{scope}

\node (stop) [end, below=of mergeDistributions,align=center,yshift=-0.7cm] {Terminate};

\draw [arrow] (chooseQMs) -- (chooseCircuits);
\draw [arrow] (chooseCircuits) -- (executeCircuits);

\draw [arrow] (computeSplitCalib) -- (prodSched);
\draw [arrow] (prodSched) -- (executeU);
\draw [arrow] (executeU) -- (mergeDistributions);


\node (calibrationBox) [draw=black!50, dashed, rounded corners, fit=(calibration), inner sep=6mm] {};
\node[fill=white] at (calibrationBox.north) {Calibration Stage};
\node (rankingBox) [draw=black!50, dashed, rounded corners, fit=(ranking), inner sep=4mm] {};
\node[fill=white] at (rankingBox.north) {Ranking};
\node (productionBox) [draw=black!50, dashed, rounded corners, fit=(production), inner sep=4mm] {};
\node[fill=white] at (productionBox.north) {Production Stage (target circuit $U$)};

\draw [arrow, line width=1.5pt] (calibrationBox.south) -- ([shift=({0,0.20})]rankingBox.north) ;
\draw [arrow, dashed, line width=1.5pt] (calibrationBox.east) -- ++(0.5,0) |- (productionBox.west);
\draw [arrow, dashed,line width=1.5pt] (rankingBox.east) -- ++(0.7,0) |- (productionBox.west);
\draw [arrow, line width=1.5pt] (productionBox.south) -- ([shift=({0,0.02})]stop.north) ;

\end{tikzpicture}
    \caption{Diagram of the strategies considered in the numerical investigation of Section~\ref{sec:numres},
    as a specific instantiation of the general strategy depicted in Fig.~\ref{fig:diagram_strategy}.}
    \label{fig:diagram_strategy_specific_numres}
\end{figure}

Here we make a numerical assessment of some of the steps described in the general protocol,
discussing first the calibration stage in Section~\ref{subsec:numres-calib}
and then focusing the analysis on the split and merge strategies in Section~\ref{subsec:numres-split-merge}. 
\new{In both cases, with respect to the general protocol shown in Figure~\ref{fig:diagram_strategy},
in this first numerical study, we consider the analysis of a \emph{specialized} protocol where most of the policies are fixed to simple rules according to Fig.~\ref{fig:diagram_strategy_specific_numres}. In particular, while in the general framework, we presented a production stage involving a schedule with multiple incremental executions and the possibility of an early stopping criterion, the policy for the schedule here is fixed to the case where the budget of total number of shots for the whole set of QPUs considered is exhausted in a single iteration. A detailed analysis of the scheduled incremental execution would involve testing many different strategies for distributing the computation \textit{in time}, while here we focus especially on optimally distributing computation among different QPUs, leaving the former analysis to future work.}

\new{Moreover, we chose not to employ any error mitigation techniques to maintain the data in its most unaltered and authentic form. This decision aligns with the primary goal of this work, which is not to propose new mitigation methods but rather to provide a clear and unbiased observation. Indeed, while it might happen that error mitigation might be more effective on a specific QPU than another, we assume that this would not change the relative unreliabilities in a relevant way. }

\subsection{Calibration and Ranking}\label{subsec:numres-calib}
\begin{figure}
    \centering
    \includegraphics[width=0.98\linewidth]{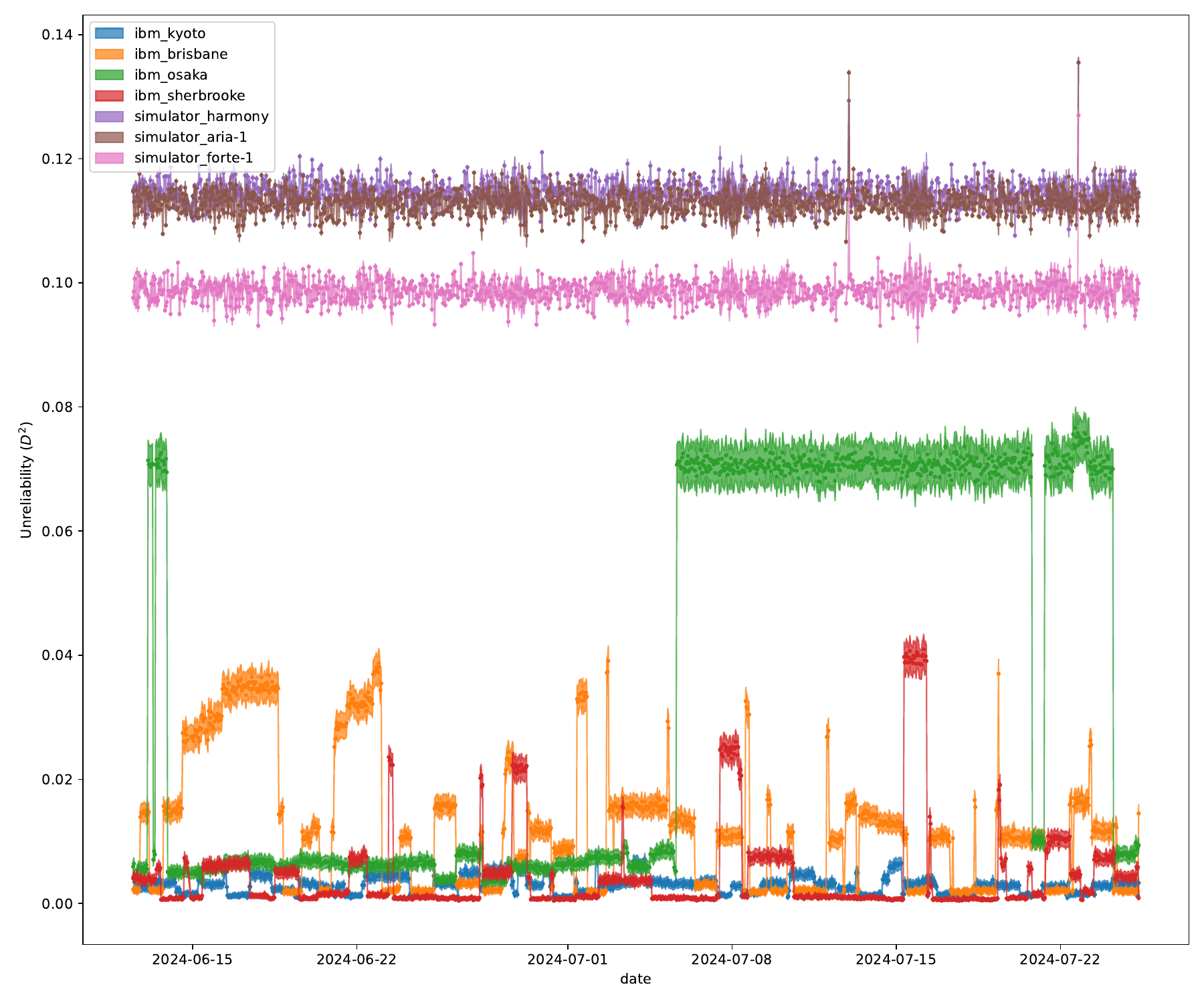}
    \caption{Behavior of the Mean Square Hellinger distance on a fixed set of $10$ random circuits $\{U_c\}$ with $q=5$ qubits as a measure of unreliability for each QPU considered in this work, reported in Table~\ref{tab:QPUs}.}
    \label{fig:history_plot}
\end{figure}

For the calibration stage, according to the diagram of Fig.~\ref{fig:diagram_strategy_specific_numres}, we first proceed with step (C.1) and select a uniform shot allocation for each QPU $m$ in the set of QPUs considered and reported in Table~\ref{tab:QPUs}, then, for the step (C.2), we select a set of 10 random circuits $\{U_c\}$ which we use as benchmark circuits~\cite{quetschlich2023mqtbench}\footnote{The OPENQAMS2 representation of the circuit is available here:\url{https://zenodo.org/records/14056270}\cite{bisicchia_dataset}}. \new{The random circuits are sampled from the unitary Haar measure.}
After execution, we choose to assign as unreliability coefficient the Mean Squared Hellinger distance of the results of each QPU (see Section~\ref{subsubsec:OptHSD}, where the performance is averaged among the set of circuits considered).
Estimates of the unreliability defined as the Mean Square Hellinger distance for each QPUs is shown in Fig.~\ref{fig:history_plot}, where measurements span a time window of about one month and a half. 
\begin{table}
    \centering
\begin{tabular}{c|c|c|c|c}
 & \multicolumn{4}{c}{unreliability} \\
QPU name & min & $25\%$ qt & median ($50\%$ qt) & $75\%$ qt \\
\cline{1-5}
ibm\_kyoto & 0.00071 & 0.0018 & 0.0029 & 0.0035 \\
ibm\_brisbane &  0.0014 & 0.0022 & 0.011 & 0.016 \\
ibm\_osaka & 0.0033 & 0.0062 & 0.0079 & 0.071 \\
ibm\_sherbrooke &  0.00044 & 0.00084 & 0.0013 & 0.0055 \\ \cline{1-5}
simulator\_harmony & 0.108 & 0.113 & 0.114 & 0.116 \\
simulator\_aria-1 & 0.107 & 0.112 & 0.113 & 0.114 \\
simulator\_forte-1 & 0.093 & 0.098 & 0.099 &0.100 \\ \cline{1-5}
\end{tabular}
\caption{Table of QPU emulators considered in this work with some statistical information such as the minimum and the $25\%$, $50\%$ (media) and $75\%$ percentiles of their unreliability.}
\label{tab:QPUs}
\end{table}
It is interesting that even if there are sensible fluctuations in time, the best performance between the QPUs considered seems always represented by `ibm\_sherbrooke' followed by `ibm\_kyoto', while it is not always clear which between `ibm\_brisbane' and `ibm\_osaka' take the second and third place in the ranking. Furthermore, according to our observation, 
the performance of IONQ QPUs appears to be one order of magnitude worse. 
Due to the variability in QPU performances, we stress the importance of a somewhat frequent calibration and assessment of the unreliabilities, at least at the order of the regions of stability (i.e., a few hours).

\subsection{Split-merge strategies}\label{subsec:numres-split-merge}
Here we follow the right side of Fig.~\ref{fig:diagram_strategy_specific_numres}, where the production
stage is simplified as a single iteration and the only variable elements are the split and weight strategies, which we test in three variants for both steps: \emph{uniform} (all shots are allocated/merged uniformly on the QPUs considered), \emph{Hellinger} (Section~\ref{subsubsec:OptHSD}) and \emph{MISE} (Section~\ref{subsubsec:OptMISE}).
For simplicity, in the split we are not including other factors that might reweight the resources allocated, for example, different cost per shot in the execution of different QPUs or in the queue and execution times (see Section~\ref{subsec:policies} for a more in depth summary of different situations or~\cite{Bisicchia2023363}).

\begin{figure}
    \centering
    \includegraphics[width=0.98\linewidth]{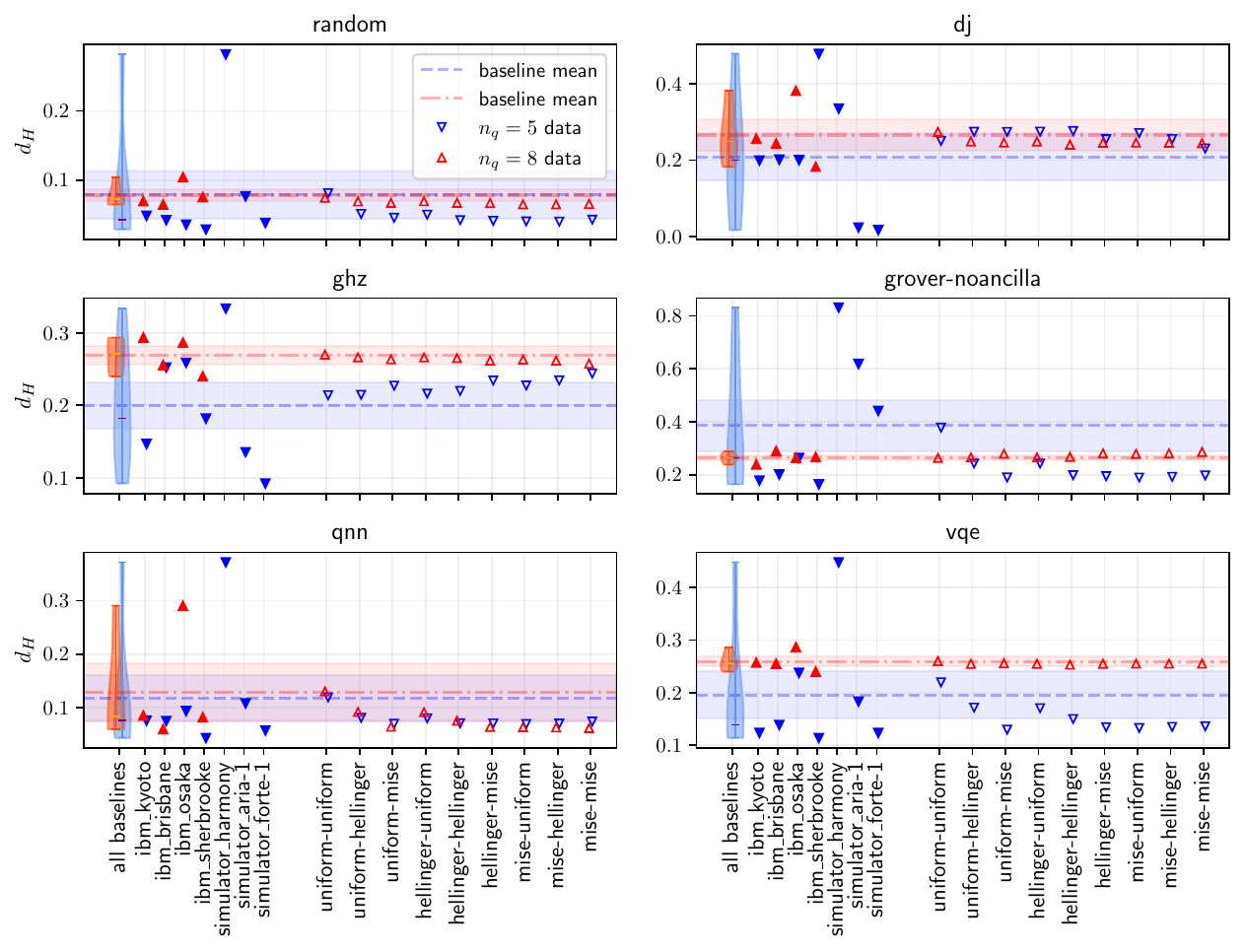}
    \caption{Results of single QPU executions (left part of the panels) and of the split and merged results from the full set of available QPUs (see text for details) in terms of the Hellinger distance ($d_H$, see Eq.~\eqref{eq:Hellinger_def}) from the ideal case. Points are slightly shifted on the horizontal axis for better readability.  
    .}
    \label{fig:numres-splitmerge_single2all}
\end{figure}

The main experimental results of this Section are shown in Fig~\ref{fig:numres-splitmerge_single2all}. Each panel represents the executions on a different circuit type, between six different cases considered. On the left part of each panel, the `baselines' are shown for each QPU, obtained by running all the shots on single QPUs (the leftmost violins refer to the overall distribution of these results among the QPUs). 
The specific benchmark circuits have been generated as qasm code for 5 and 8 qubits using the MQT Bench library~\cite{quetschlich2023mqtbench}. 
On the right part of the panels are shown data with different split and merge strategies for the set of all QPUs (ibm+ionq for the $n_q=5$ data and only IBM for the $n_q=8$ data).
We notice that, while the split-merged results never improve the best baseline for each circuit, nevertheless the former appears to be quite robust and compatible with the average between the different baselines. Furthermore, the performance of each QPU depends heavily on the circuit considered, to the extent that the trends in the performance for some circuits are not always consistent. For example, considering the GHZ circuit, results involving all the 7 QPUs considered, show the opposite trend instead of the one expected which results in an improvement from the uniform strategy to the MISE one, as observed in the other cases. This might be due to a high variance between the baselines in this case, which is not well reflected by the calibrated data which is instead trained from random circuits. 
In generality, with some exceptions as the one mentioned before, we observe that \emph{either} splitting or merging using the Hellinger or MISE strategy improves the results of just uniformly splitting and naively merging according to the uniform strategies alone. 
\new{A complete account of all different combinations of split and merge policies for each of the circuits considered is available in~\cite{bisicchia_dataset}}.

\begin{figure}
    \centering
    \includegraphics[width=0.98\linewidth]{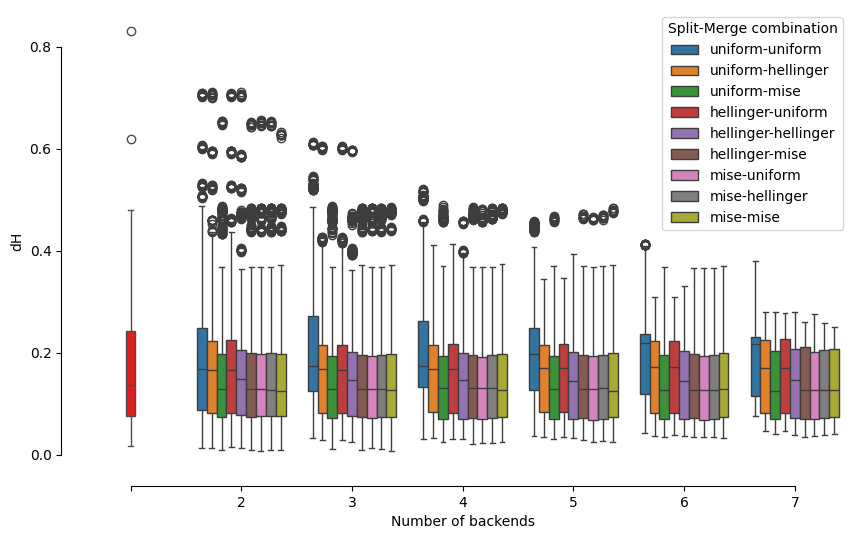}
    \caption{Measured error for all circuits, divided by split and merge policy combinations while increasing the number of available QPUs. Each bar represents a specific combination of split and merge policies. The red bar corresponds to execute all the shots on a random QPU.}
    \label{fig:allpolicies}
\end{figure}

\new{Figure~\ref{fig:allpolicies} provides a comprehensive overview of the behaviour of different split and merge policies as the number of QPUs increases. The figure shows the performance of all possible combinations of these policies under varying numbers of available QPUs. A key observation is that, as the number of QPUs increases, the maximum error consistently decreases, while the minimum error tends to rise slightly.
This behavior aligns with the underlying principle of the shot-wise methodology: typically, it is difficult for a quantum programmer to know which QPU will perform best for a given circuit at any particular time. Naturally, distributing shots across multiple QPUs can lead to a small increase in error compared to using the best-performing QPU. However, since identifying the best QPU in real-time is often impractical, this distribution approach provides a safeguard, improving the worst-case outcomes.
In fact, by applying the shot-wise distribution, the overall results tend to improve in the worst-case scenario. In the average case, the error either decreases or remains comparable to that of a single-QPU execution. Thus, this method offers a clear advantage: it consistently yields better results in the worst case, with the potential for improvement in the average case, and no significant degradation.
Additionally, we observe a reduction in the standard deviation of the results as the number of QPUs increases. This indicates that the shot-wise distribution method produces more “robust” results, with a narrower spread and a lower upper bound on error, albeit at the slight cost of potentially reduced accuracy in the best-case scenario.
}

\begin{figure}
    \centering
    \includegraphics[width=0.98\linewidth]{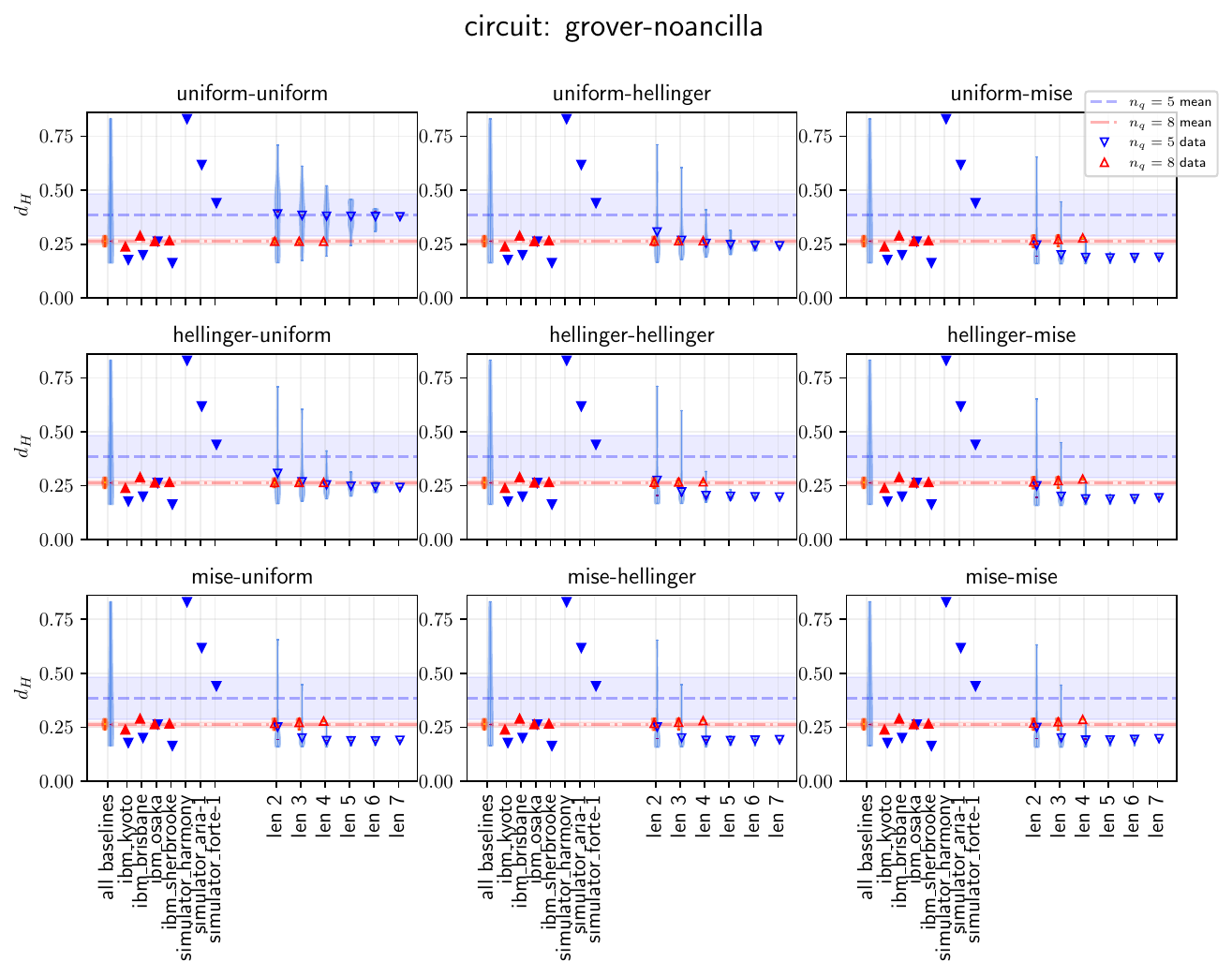}
    \caption{\new{Measured Hellinger distance ($d_H$) from ideal, for a Grover algorithm task using different split and merge policies and for different groups of up to 7 QPUs. The left side of each panel is the same and corresponds to the performance of each considered QPU as taken individually.}}
    \label{fig:grover}
\end{figure}

\begin{figure}
    \centering
    \includegraphics[width=0.98\linewidth]{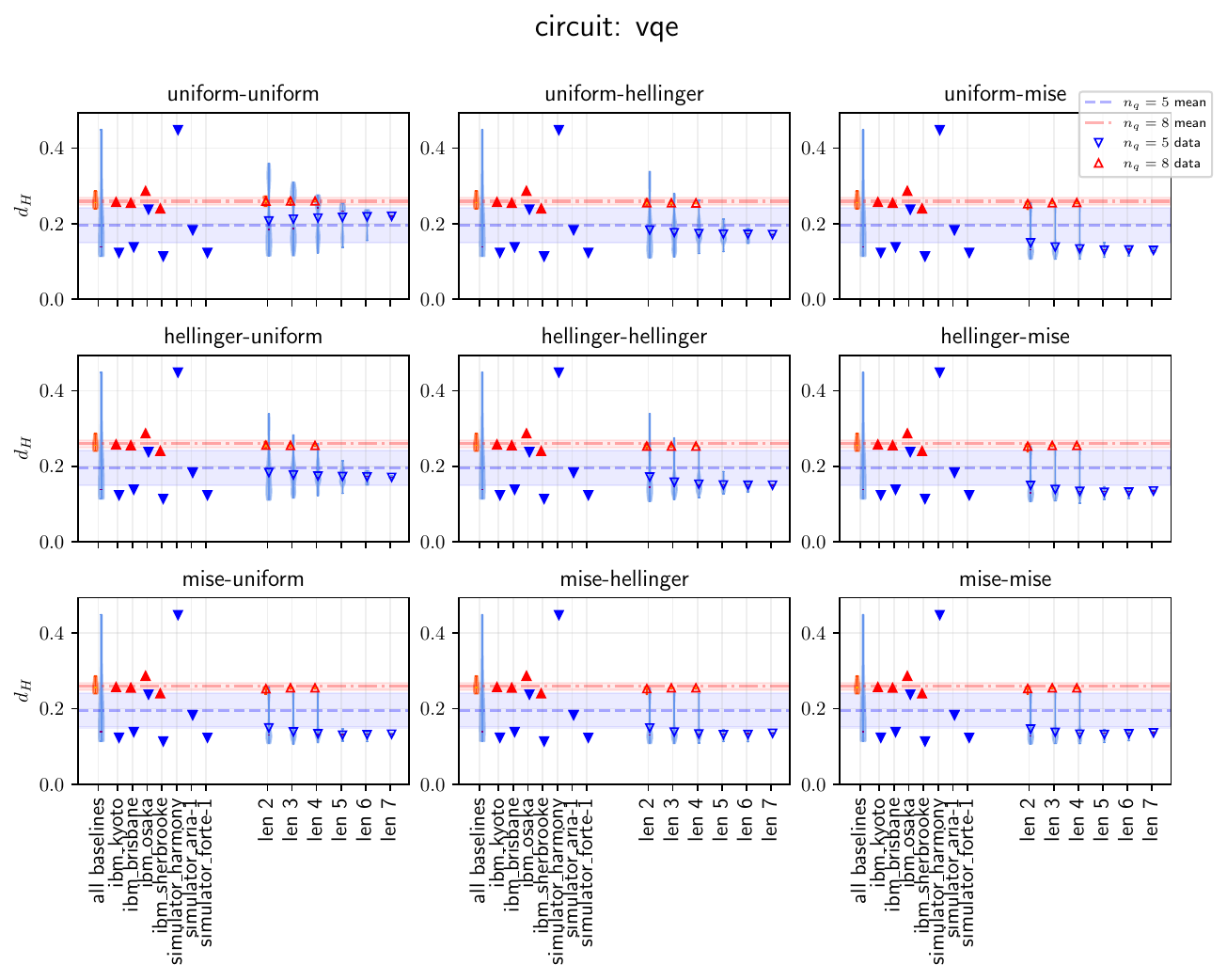}
    \caption{\new{Measured Hellinger distance ($d_H$, see Eq.~\eqref{eq:Hellinger_def}) from ideal, for a VQE task using different split and merge policies and for different groups of up to 7 QPUs. The left side of each panel is the same and corresponds to the performance of each considered QPU as taken individually.}}
    \label{fig:vqe}
\end{figure}

\new{This behaviour is, for instance, clearly illustrated in Fig.~\ref{fig:grover} and Fig.~\ref{fig:vqe}, in which for each circuit we test two different sizes (4 and 8 qubits) and all split and merge policy combinations. The results of the shot-wise distribution approach, are illustrated by distributing the shots from 2 to 7 QPUs and their results are compared with single and overall (all baselines) QPU executions.}

\subsection{Discussion}

\new{
One key advantage of shot-wise distribution in the field of Quantum Software Engineering (QSE) is its robustness to quantum noise. By distributing shots across QPUs with varying noise profiles, the impact of errors is reduced, resulting in more stable outcomes. This is crucial for QSE, as it ensures that quantum applications perform reliably across different hardware platforms. Additionally, shot-wise distribution enhances worst-case performance by mitigating the risk of relying on a single, underperforming QPU, making it useful when the best-performing QPU is uncertain.
The method also improves error resilience, as distributing shots across multiple QPUs helps balance the computational load and increases fault tolerance. If one QPU fails, others can continue the computation, ensuring continuity, which is important for long-running algorithms. Furthermore, shot-wise distribution is hardware-agnostic, allowing flexibility across different quantum architectures and making it easier to scale across a variety of platforms.
Together with these advantages, shot-wise distribution introduces some challenges. Managing multiple QPUs adds overhead in terms of coordination, scheduling, and result aggregation. This complexity can slow down execution and requires more sophisticated resource management. Additionally, it may dilute best-case performance, as spreading shots can result in lower accuracy than concentrating them on the highest-performing QPU.
Another limitation is the dependency on reliable QPU calibration. For optimal results, accurate and up-to-date information about each QPU’s performance is essential. Without it, the distribution may be inefficient. Additionally, combining results from QPUs with different error profiles requires complex aggregation techniques, adding to the computational overhead.
%
%
In conclusion, the shot-wise methodology consistently performs at least as well as the average outcome of a single quantum processing unit (QPU) while often surpassing many individual QPUs in various scenarios. This approach not only reduces variability, leading to more stable and reliable results, but also enhances overall performance when compared to executing all shots on a single QPU. Although it is not yet a complete solution for noise mitigation—a direction we intend to explore further—on average, the shot-wise method improves results and mitigates output variation. Additionally, it offers significant qualitative advantages, such as increased flexibility and customizability tailored to specific requirements. Overall, the shot-wise approach provides a stable and adaptable technique for executing shots across multiple heterogeneous quantum computers, combining both qualitative and quantitative benefits.}

\section{Related Work}\label{sec:literature}
Ever since Preskill highlighted the challenges of contemporary Quantum Computers~\cite{nisq}, researchers have designed and developed strategies to tackle or mitigate these constraints~\cite{lau2022nisq}. Within the scope of this paper — generalizable as ``approaches to perform quantum computations on NISQ devices'' — we identify, to the best of our knowledge, three primary categories:

\begin{itemize}
    \item \textit{Fighting the Noise}: this category encompasses methodologies and techniques aimed at mitigating, or ideally eliminating, the noise inherent in quantum computations. Within this realm, we discern two principal strategies:
    \begin{itemize}
        \item \textit{Quantum Strategies}: these approaches aim to combat noise \textbf{during} quantum computations by operating on the structure of the circuit to be executed.
        \item \textit{Classical Strategies}: these methods target noise \textbf{before} and/or \textbf{after} executing a quantum algorithm through pre- and post-classical processing of quantum algorithms and output distributions.
    \end{itemize}
    \item \textit{Going beyond the Intermediate scale}: here we find into methodologies seeking to execute quantum circuits larger than those achievable with NISQ devices.
    \item \textit{Distributing Quantum Computations}: this category encompasses methodologies that address the limitations of quantum computations as a whole while considering the presence of multiple heterogeneous QPUs. The objective is to optimize the execution of quantum circuits across a distributed computing environment.
\end{itemize}

While our work predominantly aligns with the latter group, it introduces, to the best of our knowledge, a novel idea: distributing \textbf{even} the same quantum circuit among multiple QPUs, exploiting the necessity to run multiple shots. The subsequent sections illustrate deeper each of these categories, presenting key related works associated with each domain.

\subsection{Fighting the Noise}
\label{sec:relatedworknoise}

The history of error correction techniques and, more broadly, strategies for combating the noise inherent in Quantum Computing is nearly as old as Quantum Computing itself.

In the early days of Quantum Computing, the excitement surrounding the field was tempered by the challenge of qubit noise. Shor, renowned for demonstrating the potential of Quantum Computing~\cite{shor1994algorithms,shor1995scheme}, injected new vitality into the field with his proposal of an initial error correction technique. This technique suggested that if the noise remained below a certain threshold, it would be possible to apply the solution and execute quantum computation as if it were devoid of noise~\cite{shor1996fault}.

As Quantum Computing progressed, error correction techniques evolved to become more sophisticated~\cite{knill1997theory,lidar2013quantum,terhal2015quantum}. However, these techniques typically demand a significantly higher number of qubits than are currently available to be effective. Consequently, while researchers continue to refine error correction techniques to be more resource-efficient, efforts have also emerged to mitigate noise while awaiting full error correction~\cite{temme2017error,endo2018practical,kandala2019error,giurgica2020digital,endo2021hybrid,cai2023quantum}.

Another compelling research direction involves the development of \textit{noise-aware compilers}. These compilers are designed to compile and optimize circuits for a given QPU while considering its topology, performance, and noise characteristics among the others. They determine crucial factors such as the initial mapping of virtual qubits onto physical qubits and the optimal set of swap operations, particularly in non-\textit{all-to-all} topologies. Such classical techniques have the potential to mitigate noise and enhance the performance of quantum circuits~\cite{heckey2015compiler,tannu2019ensemble,tannu2019not,murali2019noise,li2019tackling}.

Error correction and error mitigation techniques as well as compilation and optimization processes are all approaches working before and/or after the actual circuit execution, for that reason they are completely transparent to our \textit{shot-wise} methodology. Therefore, they are entirely compatible with our \textit{shot-by-shot} methodology and can be applied in conjunction. Furthermore, our experiments provide initial evidence that distributing circuit shots among multiple heterogeneous QPUs can reduce noise compared to executing all shots on a single QPU. Thus, we intend to explore the possibility of employing \textit{shot-wise} methodologies as error mitigation techniques and compare them with state-of-the-art approaches, potentially integrating them with already existing ones to further enhance noise reduction strategies.

\subsection{Going beyond the Intermediate scale}
\label{sec:relatedworkscale}

Achieving the capability to execute quantum circuits beyond the current hardware constraints, typically limited to a few hundred qubits at most, is essential for unlocking the full potential of quantum computing. While much attention is directed towards scaling the size of current quantum hardware, an alternative approach to circumventing this limitation involves breaking down larger circuits into smaller pieces. These fragments are then executed independently on NISQ devices, with the resulting computations merged to reconstruct the final output.

These techniques are recognized under various names, including circuit cutting, circuit knitting, and Quantum divide and conquer, or Quantum divide and compute, among others. This strategy offers a promising avenue for harnessing the computational power of existing smaller-scale quantum hardware to tackle larger quantum algorithms.

A seminal contribution is presented in~\cite{peng2020simulating}, where the authors discuss the theoretical foundations and conduct experiments on circuit cutting through tensor-network techniques. This work is further expanded upon by~\cite{ayral2020quantum,ayral2021quantum}, where the authors test the approach in the presence of noise and observe that recombining noisy fragments can outperform results without fragmentation. They also investigate the impact of different noise sources on the success of the cutting process.

Another seminal work in this field is illustrated in~\cite{cutqc}, where the authors introduce \textit{CutQC}, a scalable circuit cutting approach. They propose a method to execute quantum circuits more than twice the size of available quantum computer backends. Moreover, their approach demonstrates significant improvements in fidelity compared to direct executions on large quantum computers, along with elevated speedup over classical simulations. They utilize a mixed-integer programming approach to automate the identification of cuts requiring minimal classical postprocessing. Additionally, they discuss two types of postprocessing: full-definition (FD) query and dynamic-definition (DD) query, differing in whether the entire $2^n$ full-state probability output of the uncut circuit is reconstructed.

In a different approach, \cite{perlin2021quantum} introduces \textit{maximum-likelihood fragment tomography} to find the most likely probability distribution for the output of a quantum circuit based on measurement data obtained from circuit fragments. The core of their idea is to perform the circuit fragments by providing a variety of quantum inputs to and measuring its quantum outputs in a variety of bases. Supported by both theoretical and experimental findings they advocate for the use of circuit cutting as a standard tool for running clustered circuits on quantum hardware. Indeed, they found that circuit cutting can estimate the output of a clustered circuit with higher fidelity than full circuit execution.

A recent contribution is presented in~\cite{lowe2023fast}, where the approach is based on randomized measurements, by randomly inserting measure-and-prepare channels, to express the output state of a large circuit as a separable state across distinct devices. With this approach, they apply circuit cutting to large-scale QAOA problems on clustered graphs, i.e., up to 129-qubit problems, demonstrating the potential of circuit cutting procedures in practical applications.

In contrast to the aforementioned methodologies, our \textit{shot-wise} approach presents a significantly different perspective and can be viewed as orthogonal. While circuit cutting focuses on the \textit{circuit dimension}, our focus lies in the \textit{shot dimension}. Our strategy is agnostic to whether the shots to be distributed originate from a whole circuit or fragments. Moreover, considering that executing smaller fragments may yield better results than performing the whole circuit, we believe that combining the power of fragment cutting with the idea of executing shots from each fragment on multiple heterogeneous QPUs can further enhance performance. We intend to explore this combination in future work. An initial step in this direction is presented in~\cite{chatterjee2022qurzon}, where the authors combine circuit cutting and parallel scheduling algorithms for quantum multicomputing. However, such work still treats shots as a single monolithic entity, unlike our approach.

\subsection{Distributing Quantum Computations}
\label{sec:relatedworkdistribution}

Numerous research efforts are dedicated to identifying the most suitable Quantum Processing Unit (QPU) for executing a particular quantum circuit. These approaches typically define various metrics encompassing not only the inherent characteristics of a specific Quantum Computer (e.g., qubit count, coupling map, noise model) and the quantum circuit itself (e.g., gate count, width, depth) but also environmental factors such as queue waiting time, pricing plans, and availability periods.

One notable approach following this principle is the \textit{Quantum API Gateway}~\cite{qapigateway}. In this work, the authors devised a service that automatically selects the optimal QPU among available options for each submitted quantum circuit. The selection process takes into account factors such as QPU architecture (gate-based or annealing) and circuit width. Users have the flexibility to customize the selection criteria, specifying preferences for speed or cost efficiency.

Similarly, the \textit{NISQ Analyzer}~\cite{nisqanalyzer} employs a comparable workflow to determine the best QPU for a given quantum circuit. However, unlike the \textit{Quantum API Gateway}, this approach acknowledges that multiple circuit implementations may exist for a single quantum program. The \textit{NISQ Analyzer} automatically selects the best implementation from a repository of quantum programs and associated circuit implementations through a set of selection rules associated with each circuit implementation depending on the input data. The best QPU-circuit pair is then determined based on criteria such as circuit width, depth, and the choice of Software Development Kit (SDK).

The \textit{NISQ Analyzer} authors have also developed several extensions, including tools for comparing compiler outputs~\cite{nisqanalyzercompar}, optimizing compilation processes using Machine Learning (ML) models to discard potential compilers and Quantum Computers before compilation~\cite{nisqanalyzerml}, ranking compiled circuits for various QPUs using Multi-Criteria Decision Analysis methods~\cite{nisqanalyzerprioritization}, and optimizing these processes through ML techniques~\cite{nisqanalyzeropt}.

In contrast to these approaches, \cite{adaptive} introduces a quantum job scheduler that optimizes QPU selection by balancing estimated fidelity and expected waiting time. Conversely, \cite{qserverless} addresses the integration of Quantum Computing into classical enterprise cloud systems, selecting the most suitable single QPU based on factors like qubit count and queue length. Additionally, \cite{predictingml} presents a framework for automatically predicting the optimal combination of Quantum Computers, compilers, and compiler options for a given circuit, with a focus on maximizing fidelity in gate and measurement operations.

While existing approaches aim to identify the best QPU for a given quantum circuit, our proposal diverges from this paradigm by leveraging multiple heterogeneous Quantum Computers simultaneously. We distribute the shots even of a single quantum circuit among multiple heterogeneous QPUs. To the best of our knowledge, ours is the first proposal employing a \textit{shot-wise} distribution approach. Building upon the principles of shot-wise distribution, we have developed a prototype Quantum Service called the \textit{Quantum Broker}~\cite{bisicchia2023dispatching}, which distributes quantum computations shot-by-shot while considering user runtime requirements submitted through a Domain Specific Language (DSL). This paper advances and formalizes the concepts introduced in the Quantum Broker prototype into a unified conceptual and parametric framework. Furthermore, experimental evaluation validates the efficacy of the proposed ideas.

\section{Conclusions \& Future Work}\label{sec:concl}

In summary, our study presents a novel approach to quantum computation tailored to address the challenges posed by heterogeneous, noisy Quantum Computers. 

\new{With that aim}, we propose a methodology that advocates for a departure from the traditional monolithic execution of quantum circuits. By capitalizing on the inherent probabilistic nature of quantum mechanics, our \textit{shot-wise} approach enables distributed execution of quantum tasks across multiple \textit{noisy} Quantum Computers. We propose a general methodological framework, parameterized by a set of customizable policies, which allows for fine-grained management and distribution of shots across multiple heterogeneous \textit{noisy} Quantum Computers, even for a single quantum circuit.

Furthermore, our study introduces the concepts of \textit{calibration} and \textit{incremental execution} to further enhance the robustness and adaptability of quantum computation. \textit{Calibration} serves to pre-evaluate the reliability of different QPUs, while \textit{incremental execution} enables \textit{dynamic} resource allocation and decision-making even during the performance of a single computation.

\new{In conclusion, the shot-wise methodology emerges as a promising approach, yielding results that are at least comparable to single QPU averages and often superior to individual QPUs in diverse scenarios. This method not only reduces variability, enhancing result stability, but also improves overall performance compared to single QPU execution. While it does not fully address noise mitigation—an area for future research—the shot-wise strategy offers substantial qualitative benefits, including increased flexibility and adaptability to specific needs. Overall, it stands out as a reliable technique for executing shots across heterogeneous quantum systems, blending both qualitative and quantitative improvements.}


With this work, we aim to contribute to the development of more efficient and reliable quantum computing systems, overcoming current limitations and inspiring further research and innovation in the field.

Several interesting future research directions may emerge from this study:
\begin{description}
\item[Shot-wise Error Mitigation] In this paper, our experimental assessment provides initial evidence that the shot-wise methodology offers promise as an effective approach for managing quantum computations in the presence of diverse heterogeneous QPUs. Additionally, our findings suggest that this strategy may facilitate error cancellation of different Quantum Computers, resulting in a merged final distribution that is more reliable than the average partial distribution obtained from a QPU. However, further studies \new{and an in-depth comparison with state-of-the-art noise mitigation methodologies} are warranted to validate these preliminary findings conclusively. If confirmed, future research endeavours could focus on designing and developing robust shot-wise error mitigation techniques to enhance the effectiveness and reliability of quantum computation methodologies.
\item[Design and Development of Additional Split and Merge Policies] This paper introduced and examined four split and merge policies, embedded with and without calibration data. Future research could expand upon these policies, conducting comparative analyses to discern their efficacy. Moreover, exploring scenarios where specific combinations of policies outperform others could offer valuable insights.
\item[\new{Tailored Calibrations}] \new{In this work we presented a general, customisable framework to perform \textit{shot-wise} distribution of quantum computations. We have, then, experimentally tested the framework in a general setting in which the calibration phase is executed on random circuits to have a calibration that can be suitable in numerous scenarios. However, the \textit{shot-wise} methodology could be applied even on a specific scenario \new{(e.g., with Variation Quantum Algorithms (VQA)~\cite{vqa}) and optimize the calibration phase for that specific setting. For instance, the calibration circuits could all have the same parametric quantum circuit with random parameters when working with VQA.}}
\item[Incremental Execution, Scheduling Policies and Stop Conditions] An intriguing direction for future research involves a more comprehensive exploration of incremental execution and its impact on quantum computation performance. Accompanying this investigation, the study and development of diverse scheduling policies and stop conditions could further optimize the execution process.
\item[Additional Experiments on More Quantum Providers and Real Quantum Hardware] To further validate and strengthen the findings of this study, conducting additional experiments involving multiple quantum providers and utilizing real quantum hardware is essential. By exploring various environments and defining practical benchmark use cases, researchers may gain deeper insights into the robustness and applicability of proposed strategies.
\end{description}

\acknowledgments
GC and MD acknowledge support from Fondazione ICSC - National Centre on HPC, Big Data and Quantum Computing - SPOKE 10 (Quantum Computing) and received funding from the European Union Next-GenerationEU - National Recovery and Resilience Plan (NRRP) – MISSION 4 COMPONENT 2, INVESTMENT N. 1.4 – CUP N. I53C22000690001. JGA and JMM were partially funded by the European Union “Next GenerationEU /PRTR”, by the Ministry of Science, Innovation and Universities (TED2021-130913B-I00, RED2022-134148-T, and PDC2022-133465-I00). They were also supported by QSERV project funded by the Spanish Ministry of Science and Innovation and ERDF; by the Regional Ministry of Economy, Science and Digital Agenda of the Regional Government of Extremadura (GR21133); and by European Union under the Agreement - 101083667 of the Project “TECH4E -Tech4effiency EDlH” regarding the Call: DIGITAL-2021-EDlH-01 supported by the European Commission through the Digital Europe Program.

\bibliographystyle{ACM-Reference-Format}
\bibliography{refs}

\appendix

\section{Optimal distributions and weights}\label{sec:weights_opt}
In this Section we show some details about the optimality criteria and the 
merged probability distributions for the Weighted Square Hellinger Distance (Section~\ref{sec:WSHD_weights_opt}) and the Mean Integrated Square Error (Section~\ref{sec:MISE_opt}),
as well as the optimal weights for the split.

\subsection{Optimal Weighted Square Hellinger Distance}\label{sec:WSHD_weights_opt}
In the case of $d_H$ being the Hellinger distance and denoting by $(m)$ the quantity associated to
the $m$-th QPU in the set of QPUs considered, the expression in Eq.~\eqref{eq:DSquare} becomes
\begin{align}
D^2_{\text{Hell}}(\bar{p};p^{(m)},w^{(m)}) = 1-\sum_{m=0}^{M-1} w^{(m)} \sum_x\sqrt{\bar{p}_x p^{(m)}_x}.
\end{align}
For fixed weights $w^{(m)}$, it is straightforward to show that the optimal solution, which minimizes $D^2_{\text{Hell}}$ with constraints $0\leq \bar{p}_x \leq 1\;\forall x$, is
\begin{align}
\bar{p}^{* (\text{Hell})}_x = {\Bigg[\sum_{m=0}^{M-1} w^{(m)} \sqrt{p^{(m)}_x}\Bigg]}^2.
\end{align}
As before, since we do not have direct access to the actual QPU
distributions $p^{(m)}$, but only to their relative counts $\frac{\hat{c}^{(m)}}{n^{(m)}}$, we must take into account the bias due to 
the non-linearity of the square root, as discussed in Appendix~\ref{sec:biasremoval}.

\subsection{Optimal Mean Integrated Square Error}\label{sec:MISE_opt}
It is useful to formally decompose the MISE in
Eq.~\eqref{eq:MISEdef} as a sum of two contributions
\begin{align}
    \textrm{MISE}(\vec{w};\vec{n}) = \textrm{VAR}(\vec{w};\vec{n}) +\textrm{BIAS}^2(\vec{w}),
\end{align}
defined as
\begin{align}
    \textrm{VAR}(\vec{w};\vec{n}) &\equiv \sum_x\mathbb{E}_{\mathcal{D}=\cup_{m} \mathcal{D}^{(m)}} \Big[ {(\hat{p}_x^{(\vec{w};\vec{n})}[\mathcal{D}] - p^{(\vec{w})}_{x})}^2 \Big],\\
    \textrm{BIAS}^2(\vec{w}) &\equiv \sum_x {(p_x^{(\vec{w})} - p^{(\text{ideal})}_{x})}^2,
\end{align}
where the first encodes the fluctuations for
different realizations of the dataset $\mathcal{D}$ around $p^{(m)}_x$, while the second, 
independent from the dataset, quantifies
between the target distribution $p^{(ideal)}_x$ and
the one obtained from a convex weighted average.

While one cannot compute exactly $\mathrm{VAR}$ and
$\mathrm{BIAS}$, we can nevertheless estimate them 
from resamples of a single realization of a dataset.
For example, even without knowing the exact weighted
distribution $p_x^{(\vec{w})}$, we can estimate
$\mathrm{VAR}$ from two datasets $\mathcal{D}$ and $\mathcal{D}'$, sampled independently from as a multinomial with $p_x^{(\vec{w})}$, as
\begin{align}
    \textrm{VAR}(\vec{w};\vec{n}) &\simeq \frac{1}{2} \sum_x {(\hat{p}_x^{(\vec{w};\vec{n})}[\mathcal{D}] - \hat{p}_x^{(\vec{w};\vec{n})}[\mathcal{D}'])}^2,
\end{align}
or, more practically, as an average between many 
bootstrap resamples of a single dataset.

Assuming we can always find a convex solution 
in the bulk of the simplex made of weight parameters $w_m \in [0,1]$ and $\sum_{m=0}^{M-1} w_m =1$,
we can minimize the MISE, including 
the normalization constraint on the weights, 
by adding a Lagrange multiplier $\mu$ as follows
\begin{align}
\Lambda(\vec{w},\mu;\vec{n}) &\equiv \textrm{MISE}(\vec{w};\vec{n}) + 2 \mu (\sum_m w_m -1).
\end{align}
The function $\Lambda$ can then be minimized 
as customary by computing the derivatives 
with respect to $\vec{w}$ and $\mu$, which
results in a linear system $C \vec{w} = \vec{f}-\mu\vec{1}$ where
\begin{align}
C_{m,m'} &\equiv \sum_x \mathbb{E}_{\mathcal{D}}[ \hat{p}_x^{(m;n_m)}[\mathcal{D}]\hat{p}_x^{(m';n_{m'})}[\mathcal{D}]],\\
f_m &\equiv \sum_x p^{(m)}_x p^{(ideal)}_x. 
\end{align}
Therefore, tuning $\mu$ in such a way to make $w_k$
properly normalized (enforcing 
$\partial_{\mu} \Lambda = 0$), we have
\begin{align}\label{eq:MISEopt_wmu}
\bar{w}_m &= \sum_{m'} {({C}^{-1})}_{m,m'} \Big[f_{m'} - \bar{\mu}\Big],\\
\bar{\mu} &\equiv \frac{\Big(\sum_{m',m}{({C}^{-1})}_{m',m} f_m -1\Big)}{\sum_{\tilde{m}',\tilde{m}}({C}^{-1})_{\tilde{m}',\tilde{m}}}.
\end{align}
It might happen that some values of $\bar{w}_m$
are negative and cannot be interpreted as convex
weighted average in the bulk of the weight simplex. 
In that case, one can exclude QPUs $m$ with $\bar{w}_m <0$ and compute the analysis on the remaining subset of QPUs.
In general, the exact QPU distributions
$p^{(m)}_x$ are not available, 
so the matrix $C$ and vector $f$ have to be estimated
from the respective counted distributions,
as well as $\bar{w}_m$ and $\bar{\mu}$, whose bias
should be possibly removed via resampling techniques.

For the calibration stage, we consider a number $N_c>1$ of circuits, 
associated with ideal distributions $p_x^{(ideal,c)}$.
The optimal weights can still be estimated from Eq.~\eqref{eq:MISEopt_wmu}, 
but where the matrix $C$ and vector $f$ come from a minimization of the 
average $MISE^{(c)}$ for each circuit $c$ and weighted according to the relative number of shots per circuit $\frac{n^{(c)}}{n_{\text{tot}}}$, which results in
\begin{align}
C_{m,m'} &\equiv \sum_{c=0}^{N_c-1} \frac{n^{(c)}}{n_{\text{tot}}}  \mathbb{E}_{\mathcal{D}^{(c)}}\Big[\sum_{x} \big(\hat{p}_x^{(m,c;n_m)}[\mathcal{D}^{(c)}]\hat{p}_x^{(m',c;n_{m'})}[\mathcal{D}^{(c)}]\big)\Big],\\
f_m &\equiv \sum_{c=0}^{N_c-1} \frac{n^{(c)}}{n_{\text{tot}}} \sum_x p^{(m,c)}_x p^{(ideal,c)}_x,
\end{align}
where $\mathcal{D}^{(c)}\equiv \cup_{m} \mathcal{D}^{(m,c)}$ denotes
the union of the datasets from each QPU for circuit $c$,
while the number of shots per circuit is 
$n^{(c)}=\sum_{m} n^{(m,c)}$.

\section{Removing the bias in distance estimates}\label{sec:biasremoval}
Let us consider the estimation of the Hellinger distance between
a distribution $p$ sampled with $n$ shots, and a known distribution $q$.
The dataset of the sampled distribution can be written as $\{y_i\}_{i=0}^{n-1}$, and the counts read $\hat{c}_x = \sum_{i=0}^{n-1} \delta_{x,y_i}$.
A naive estimate of the Hellinger distance between $p$ (unknown) 
and $q$ (known) reads
\begin{align}
    \hat{d}_{H}\equiv d_{H}(\hat{p},q) = \sqrt{1-\sum_x\sqrt{\frac{\hat{c}_x}{n} q_x}}.
\end{align}
However, as mentioned in Sec.~\ref{sec:methods}, this typically overestimates
the true distance, since $\mathbb{E}[d_H^2(\hat{p},q)]\geq d_H(p,q)$ by Jensen inequality,
with equality holding only in boundary cases.
The jackknife estimate of the Hellinger distance is defined as
\begin{align}
    \hat{d}_{H {\text{jack}}}\equiv \frac{1}{n}\sum_{i=0}^{n-1} \hat{d}_{H (i)},
\end{align}
where $\hat{d}_{H (i)}$ is the naive estimate of $d_H$ computed
by removing the $i$-th count from the dataset. It is straightforward to prove that, in the case of a multinomial distribution, the 
jackknife estimate of the Hellinger distance reads
\begin{align}\label{eq:Hjack}
    \hat{d}_{H \text{jack}} &= \frac{1}{n}\sum_{y}\hat{c}_y \hat{d}_{H (y)},\\
    \Delta d_{H \text{jack}} &= \sqrt{\frac{n-1}{n}\sum_{y} \hat{c}_y 
    {\Big(\hat{d}_{H (y)} -\hat{d}_{H \text{jack}}\Big)}^2}
\end{align}
where
\begin{align}
    \hat{d}_{H (y)}\equiv \sqrt{1-\frac{1}{\sqrt{n-1}}\Big[\sqrt{n}(1-\hat{d}_H^2) - \Big(\sqrt{\hat{c}_y}-\sqrt{\hat{c}_y-1}\Big)\sqrt{q_y}\Big]}.
\end{align}
The estimate in Eq.~\eqref{eq:Hjack} is still affected by bias $O(n^{-1})$, 
which can nevertheless be easily corrected with the following 
prescription~\cite{cameron2005microeconometrics}
\begin{align}\label{eq:Hellinger_biasred}
    \hat{d}^*_{H \text{jack}} = \hat{d}_{H \text{jack}} + n(\hat{d}_H-\hat{d}_{H \text{jack}}).
\end{align}
Finally, the case of a Hellinger distance where both probability 
distributions are estimated through two separate datasets, 
the analysis proceeds as before, by taking into consideration the 
independence between the extractions for the two datasets involved.

\end{document}